**Xenon isotopes identify large scale nucleosynthetic heterogeneities across the solar system**

*short title: Xenon isotopes in the solar system*

G. Avice[1*], M. Moreira[1], J. D. Gilmour[2]

*corresponding author's email: avice@ipgp.fr

[1]Université de Paris, Institut de physique du globe de Paris, CNRS, F-75005 Paris, France

[2]Dept of Earth and Environmental Science, School of Natural Sciences, University of Manchester, Manchester M13 9PL, United Kingdom

**Abstract**

Nucleosynthetic isotopic anomalies in meteorites and planetary objects contribute to our understanding of the formation of the solar system. Isotope systematics of chondrites demonstrate the existence of a physical separation between isotopic reservoirs in the solar system.

The isotopic composition of atmospheric xenon (Xe) indicates that its progenitor, U-Xe, is depleted in $^{134}$Xe and $^{136}$Xe isotopes relative to Solar or Chondritic end-members. This deficit supports the view that nucleosynthetic heterogeneities persisted during the Solar System formation. Measurements of xenon emitted from comet 67P/Churyumov-Gerasimenko (67P) identified a similar, but more extreme, deficit of cometary gas in these isotopes relative to Solar gas.

Here we show that the data from 67P demonstrate that two distinct sources contributed xenon isotopes associated with the r-process to the solar system. The h-process contributed at least 29% (2σ) of solar system $^{136}$Xe. Mixtures of these r-process components and the s-process that match the heavy isotope signature of cometary Xe lead to depletions of the precursor of atmospheric Xe in p-only isotopes. Only the addition of pure p-process Xe to the isotopic mixture brings $^{124}$Xe/$^{132}$Xe and $^{126}$Xe/$^{132}$Xe ratios back to solar-like values. No pure p-process Xe has been detected in solar system material, and variation in p-process Xe isotopes is always correlated with variation in r-process Xe isotopes. In the solar system, p-process incorporation from the interstellar medium happened before incorporation of r-process nuclides or material in the outer edge of the solar system carries a different mixture of presolar sources as have been preserved in parent bodies.

## 1. INTRODUCTION

The series of steps leading to the formation of the solar system efficiently mixed material throughout the entire protoplanetary disk *via* large scale dust drift and orbital perturbations of planetesimals leading to collisions. However, some heterogeneities remain and are detected by remote sensing of asteroids (*e.g.* (Alexander et al. 2018)), by analyses of meteorites, the remnants of the formation of the solar system, and by analyses of planetary material. During the last decade, studies have demonstrated the existence of an isotopic dichotomy (*e.g.* for Cr, Ti, Mo, Ni) between carbonaceous (CC) and non-carbonaceous (NC) chondritic meteorites (Burkhardt et al. 2019; Kruijer et al. 2017; Warren 2011). For most chemical elements, the isotopic differences between reservoirs are very small, on the order of parts per 10,000 or even lower. This dichotomy implies that the parent bodies of CC and NC meteorites incorporated different populations of presolar grains leading to distinct isotopic reservoirs still present in the modern solar system. The current favored hypothesis for establishing this dichotomy is the formation of Jupiter, which acted as a physical barrier preventing mixing of grains between outer (CC-like) and inner (NC-like) regions of the disk (Kruijer et al. 2017). The isotopic reservoir located in the outer solar system (CC-like material) carries excesses of neutron-rich Cr, Ti, Mo and Ni isotopes created in neutron-rich stellar environments (Nanne et al. 2019).

Noble gases are tracers of choice for understanding nucleosynthetic heterogeneities in solar system material (Reynolds 1960). Studies identified several nucleosynthetic end-members for noble gases (reviewed by (Ott 2014)), and attempted to establish a link between these end-members and presolar grains (SiC, nanodiamonds, etc.) (Lewis et al. 1994). Xenon, the heaviest noble gas, is particularly useful to understand the distribution of nucleosynthetic processes in the Solar System. Its nine isotopes are produced *via* different combinations of nucleosynthetic pathways: p-process only for $^{124}$Xe and $^{126}$Xe, p- and s-processes for $^{128}$Xe and $^{130}$Xe, p-, s- and r-processes for $^{129}$Xe, s- and r- for $^{131}$Xe and $^{132}$Xe and r-process only for $^{134}$Xe and $^{136}$Xe. Establishing the isotopic inventory of the p-process only isotopes $^{124}$Xe and $^{126}$Xe in the solar system is challenging since they are the rarest Xe isotopes (about 0.2% of total Xe), which makes them difficult to measure precisely. Spallation processes in space or on planetary surfaces also lead to excesses in these isotopes and corrections rely on poorly defined production rates (*e.g.* Avice et al. 2018a) for correction attempts on Martian Xe trapped in the Martian meteorite Tissint). These difficulties prevent using the distribution of light Xe isotopes to put constraints on existing nucleosynthetic models (Arnould & Goriely 2003).

One important component revealed by analyses of Xe in meteoritic nanodiamonds is Xe-HL (Lewis et al. 1987). This Xe component is characterized by extreme enrichments, sometimes higher than 100%, in both heavy ($^{134,136}$Xe) and light ($^{124-128}$Xe) isotopes relative to the solar composition. This systematic isotopic pattern is striking since there is no known nucleosynthetic process able to produce light and heavy Xe isotopes at the same time (Arnould & Goriely 2003). One alternative explanation invokes a separation process of radioactive precursors of Xe before the end of the nucleosynthesis event (Ott 1996). Gilmour & Turner (2007) used multi-dimensional correlations to reveal that subtle variations in relative enrichments of heavy ($^{134,136}$Xe) and light ($^{124-128}$Xe) isotopes do exist. This demonstrates that their parent nucleosynthetic processes can be distinct.

Xenon in the Earth's atmosphere is distinct from cosmochemical end-members. It is enriched in heavy isotopes and depleted in light isotopes by 3-4 % per atomic mass unit relative to cosmochemical end-members (Chondritic Xe or Solar Xe, Fig. 1). This mass-dependent fractionation is likely due to escape of Xe from the atmosphere until the end of the Archean eon (Avice et al. 2017; 2018b; Bekaert et al. 2018; Pujol et al. 2009; Zahnle et al. 2019). After correction for the mass-dependent isotopic fractionation, the precursor of atmospheric Xe presents selective depletions of $^{134}$Xe and $^{136}$Xe relative to $^{132}$Xe and to the isotopic composition of xenon in the Solar Wind. This observation led to the definition of a theoretical starting isotopic composition for atmospheric Xe labeled U-Xe ((Pepin 1991), Table 1). This component is solar-like (*i.e.* identical to SW-Xe) for all Xe isotopes except $^{134}$Xe and $^{136}$Xe, which are depleted by 5 and 9 %, respectively, relative to SW-Xe. Until recently, no samples containing Xe with a clear affinity with U-Xe had been found except for one study of the Tatahouine diogenite (Michel & Eugster 1994), not confirmed in a following investigation (Busemann & Eugster 2002). One study of Xe trapped in lunar anorthosites (Bekaert et al. 2017) also identified potential hints of U-Xe on top of terrestrial-like Xe isotope composition but measurements of Xe in other lunar samples do not show this tendency (Mathew & Marti 2019).

The ROSINA instrument onboard the Rosetta spacecraft measured the isotopic composition of gases in the coma of comet 67P/Churyumov-Gerasimenko. Compared to solar Xe (SW-Xe; Meshik et al. 2014), cometary xenon (67P/CG-Xe hereafter; Marty et al. 2017) presents a solar-like pattern for $^{128,130,131}$Xe/$^{132}$Xe but an excess of $^{129}$Xe (linked to the radioactive decay of $^{129}$I) and, of interest here, a depletion of $^{134}$Xe and $^{136}$Xe (Fig. 1). U-Xe, the progenitor of atmospheric Xe, would thus consist of a mix of about 78% chondritic Xe with about 22% cometary Xe. Given the high noble gas to water ratio of comets (Balsiger et al. 2015), a cometary contribution to atmospheric Xe remains compatible with the isotopic budget of water which is close to the chondritic component (Marty et al. 2016). The ROSINA instrument onboard the Rosetta spacecraft also measured the isotopic composition of krypton emitted by

67P/Churyumov-Gerasimenko (Rubin et al. 2018). Contrary to the case of Xe, the isotopic composition of Kr is close to the solar values. In summary, the isotopic composition of the progenitor of atmospheric Xe shows depletions in r-process Xe isotopes and cometary Xe seems to share this feature.

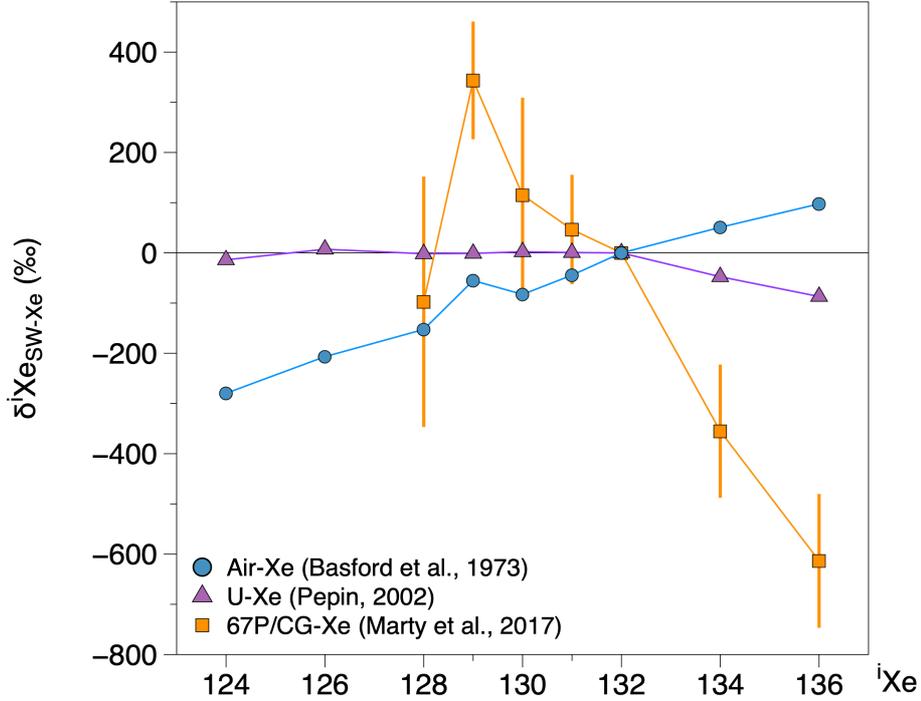

**Figure 1:** Isotopic compositions of Xe in the Earth's atmosphere (Basford et al. 1973), in the coma of comet 67P/C-G (Marty et al. 2017) and of the U-Xe component (Pepin 1991), the theoretical progenitor of Earth's atmospheric Xe. Isotope ratios of the different components are expressed relative to $^{132}$Xe, to the isotopic composition of Xe in the Solar Wind (SW-Xe) and using the delta notation: ($\delta^i$Xe$_{SW-Xe}$=(($^i$Xe/$^{132}$Xe)$_{component}$/($^i$Xe/$^{132}$Xe)$_{SW-Xe}$-1)x1000). Errors at 1σ. For Air-Xe and U-Xe, error bars are smaller than the points.

In this study, we revisit the interpretation proposed by Marty et al. (2017) and demonstrate that the nucleosynthetic mixture explaining 67P/CG-Xe leads to a severe depletion of light Xe isotopes ($^{124-128}$Xe) of the progenitor of atmospheric Xe relative to the solar composition. This depletion cannot be simply compensated by adding a contribution from HL-Xe to the nucleosynthetic mixture since it also carries r-process Xe isotopes ($^{134}$Xe and $^{136}$Xe). This requires the existence of a pure p-process Xe component that can bring $^{124-128}$Xe/$^{132}$Xe ratios back to solar-like values. This observation implies that large-scale p-process/r-process nucleosynthetic heterogeneities persisted since the formation of the solar system.

**2. OBSERVATIONS AND MODEL**

The isotopic compositions of Xe components used, and determined, in this study are listed in Table 1. Following Marty et al. (2017), the isotopic composition of Xe in the coma of 67P/C-G is taken from their period (b) of measurement. Taking values from the other two periods instead does not change the results of the present study. The composition of U-Xe is from (Pepin 1991) and of Air-Xe from Basford et al. (1973). Spinel-Xe corresponds to the isotopic composition computed by Ballad et al. (1979) for Xe in Allende spinels. In the present study isotopic ratios of Xe are normalized, illustrated (except for Fig. 2 where the normalizing composition is 67P/CG-Xe) and discussed relative to a reference value, which is the composition of the Solar Wind measured by the Genesis mission (SW-Xe, Meshik et al. 2014).

Conventionally, nucleosynthesis of the heavy isotopes of xenon involves contributions from the S-process and from the R-process (Capitalization is used to distinguish these expected components from those discussed below). Measurements of the S-process from presolar SiC grains agree well with theoretical predictions (Lewis et al. 1994). The R-process composition can be calculated by subtracting this S-process composition from the composition of solar xenon on the basis that $^{130}$Xe is produced only in the S-process.

However, from a compilation of literature analyses of presolar material (nanodiamonds and SiC grains), Gilmour and Turner (2007) identified that three, not two, major processes were required to account for the observed variation for heavy Xe isotopes. They denoted these s, r and h. Based on fits of planes with $^{130}$Xe/$^{132}$Xe and $^{136}$Xe/$^{132}$Xe as the independent variables and the ratio of each other xenon isotope to $^{132}$Xe as the dependent variable, they calculated ranges of allowed compositions for each component. They noted that the correlation extended to the light, p-process isotopes, though subtle variation was present. Gilmour and Turner's s-process, defined as making no contribution to $^{136}$Xe, was consistent with the S-process identified by Lewis et al. (1994). The remaining variation could be accounted for by two processes that each made no contribution to $^{130}$Xe and so could be seen as variants of the R-process. Each had a range of possible compositions. One, "h", accounted for the enrichment in $^{134}$Xe and $^{136}$Xe observed in presolar nanodiamonds. The other, "r", was consistent with a composition lacking $^{128}$Xe, $^{130}$Xe and $^{136}$Xe.

The relative proportions of h and r that contribute to solar xenon depend on the compositions adopted; if the r-process composition is identical to R-process xenon (which is within the allowed range for r), there is no contribution from the h process to solar xenon and the presolar material on which the analysis was based simply samples a very rare nucleosynthetic site. However, this alternative makes it coincidental that the mixing among the components yields an r-process composition producing no $^{136}$Xe. For this reason, Gilmour and Turner considered that all three processes made a significant contribution to solar system Xenon, in which case up to 31% of the solar system $^{132}$Xe budget (and all of the $^{136}$Xe) may come from the h-process.

Marty et al. (2017) reconstructed the best fit planes Gilmour and Turner used to constrain endmember compositions by considering mixing among their s-process and the two extreme compositions identified for the r-process. They did not include any h-process composition. This has the disadvantage of treating uncertainties on the endmembers, which arise from underlying uncertainties on the fitted planes and so are correlated, as independent errors. Marty et al. showed that a "mixture" of 34% of s-process, 41% of one candidate r-process ("r1") composition and 25% of the other candidate r2-process ("r2") reproduces 67P/CG-Xe. This demonstrated that the composition measured in 67P/CG lies on the planes identified by Gilmour and Turner, and lies between the endmember candidate compositions for the r-process. However, the percentage attributed to each of the alternates r1 and r2 has no physical meaning. It thus reduces the range of possible r-process compositions such that 0 < $^{136}$Xe/$^{132}$Xe < 0.25 for the r-process, the upper limit corresponding to the ratio of $^{136}$Xe/$^{132}$Xe in 67P once $^{132}$Xe has been corrected for a contribution to $^{132}$Xe from the s-process: 0.188 ± 0.068 (1σ). That is to say, the upper limit presumes that xenon from 67P is a binary mixture of the s- and r-processes. This new constraint on the composition of the r-process Xe requires that the h-process made a significant contribution to solar system xenon — at least 29% (2σ) of $^{136}$Xe came from this process.

In order to predict the light isotope composition corresponding to the heavy xenon isotopes measured in 67P, we compute a theoretical isotopic composition for cometary Xe labeled Comet-Xe following equation (1).

$$\frac{^{i}Xe}{^{132}Xe} = A \frac{^{130}Xe}{^{132}Xe} + B \frac{^{136}Xe}{^{132}Xe} + C \quad (i = 124,126,128,129,131,134).$$

(1)

$^{130}$Xe/$^{132}$Xe and $^{136}$Xe/$^{132}$Xe are the measured $^{130}$Xe/$^{132}$Xe and $^{136}$Xe/$^{132}$Xe ratios of 67P/CG-Xe (Marty et al. 2017). A, B and C are the parameters of the fitted planes presented by Gilmour and Turner (2007) in their Table 1. Errors were propagated by using a Monte Carlo algorithm. Mixing 22% of this synthetic cometary component with 78% of chondritic (Q-Xe) (Busemann et al. 2000; Marty et al. 2017) leads to a theoretical progenitor of Earth's atmospheric Xe labeled Start-Xe. Comet-Xe and Start-Xe are listed in Table 1.

## 3. RESULTS

Our theoretical cometary Xe (Comet-Xe, Table 1) has an isotopic composition closely matching 67P/CG-Xe for $^{128,130-136}$Xe (Fig. 2) but distinct from 67P/CG-Xe and SW-Xe for $^{124,126}$Xe and $^{129}$Xe (Fig. 3).

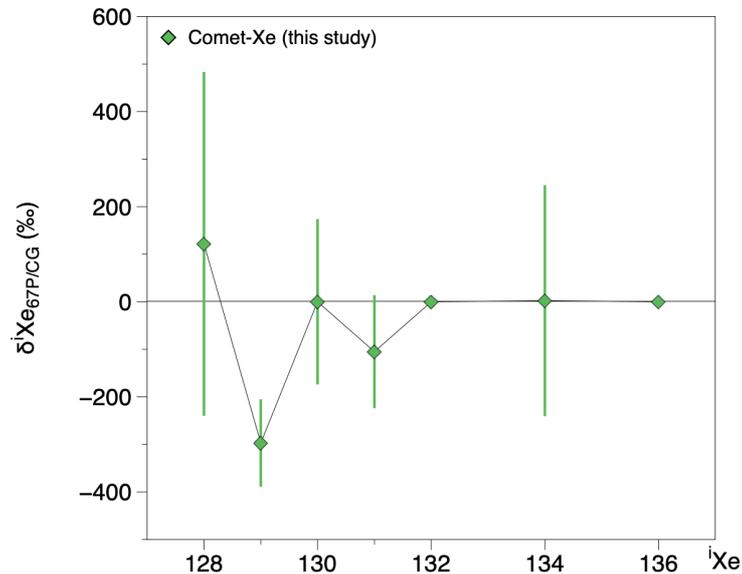

**Figure 2: Isotopic composition of Comet-Xe determined in this study and compared to 67P/CG-Xe.** Isotope ratios are expressed following the same rationale as in Fig. 1 but with 67P/CG-Xe (Marty et al, 2017) as the reference value. Errors at 1σ.

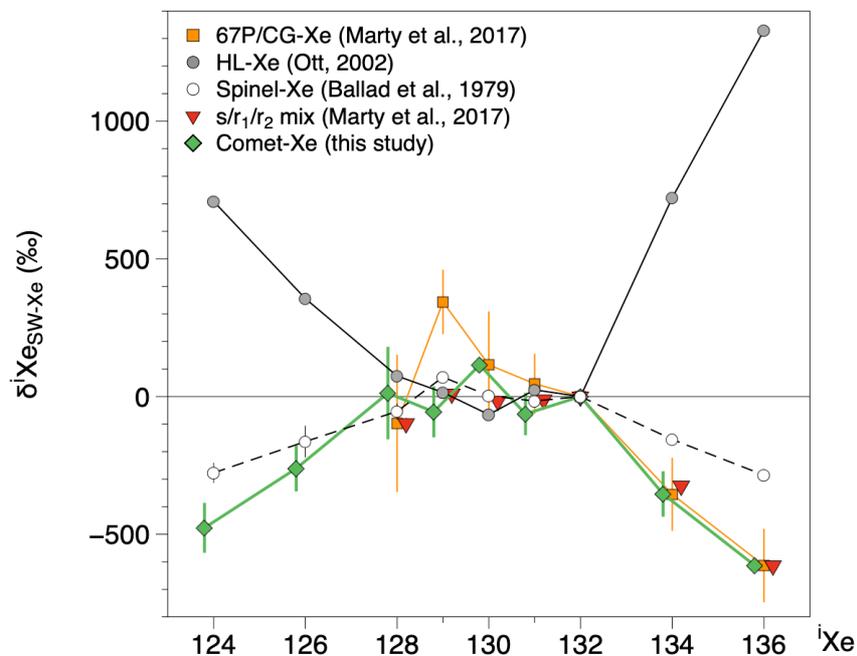

**Figure 3:** Isotopic composition of cometary Xe extended to $^{124}$Xe and $^{126}$Xe. Isotopic compositions of Xe in the Earth's atmosphere (Basford et al. 1973), in the coma of comet 67P/C-G (Marty et al. 2017), of the U-Xe component, the theoretical progenitor of Earth's atmospheric Xe and of Xe in Allende spinels

(Ballad et al. 1979) are also shown. Isotope ratios are expressed following the same rationale as in Fig. 1. Errors at 1σ.

When 22% of Comet-Xe is mixed with 78% of Q-Xe in order to reproduce the starting isotopic composition of atmospheric Xe (Marty et al. 2017), the resulting isotopic composition (labeled Start-Xe here) is also distinct from U-Xe for $^{124}$Xe, $^{126}$Xe and $^{128}$Xe (Fig. 4). The depletion relative to SW-Xe reaches -162±31, -77±30 and -16±37 ‰ for $^{124}$Xe, $^{126}$Xe and $^{128}$Xe, respectively.

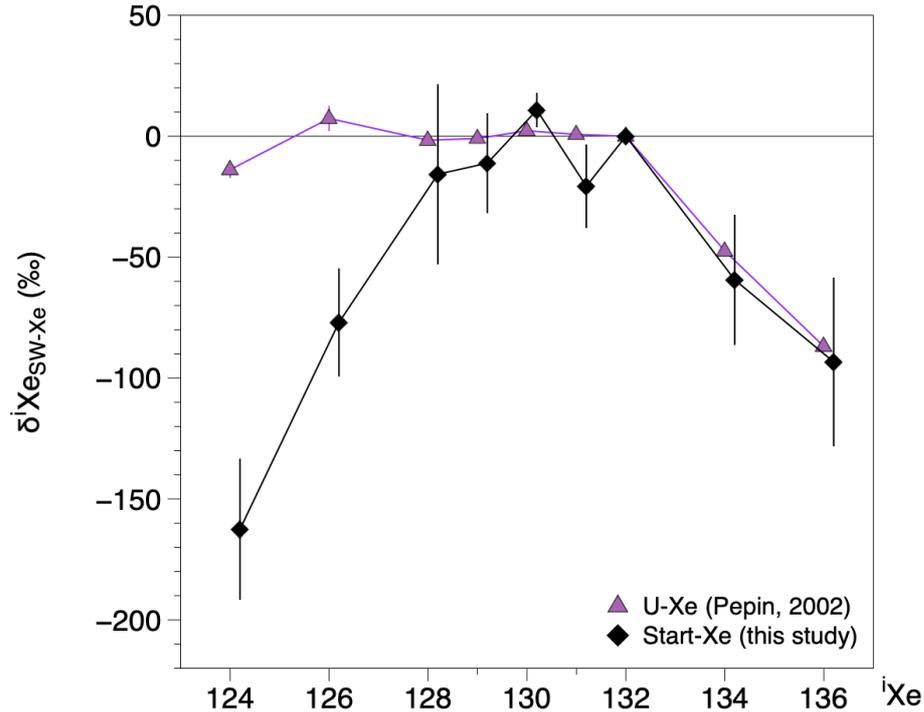

**Figure 4: Isotopic composition of Xe in U-Xe (Pepin 2002) and in Start-Xe determined in this study.** Note the difference in scale for the y-axis compared to Fig. 1-3. Isotope ratios are expressed following the same rationale as in Fig. 1. Errors at 1σ.

### 4. INTERPRETATIONS

The depletion in $^{129}$Xe of Comet-Xe relative to 67P/CG-Xe (Fig. 2) might be due to the presence of nucleosynthetic $^{129}$Xe in cometary material (Marty et al. 2017). Marty et al. (2017) highlighted that the $^{129}$Xe excess measured in 67P/CG is difficult to explain solely by radiogenic ingrowth of $^{129}$Xe coming from the decay of extinct $^{129}$I ($T_{1/2}$ = 16 Ma) since it requires orders of magnitude too much iodine relative to Xe compared to a solar-like elemental ratio. However, s-process $^{129}$Xe is accounted for in the model and r-process $^{129}$Xe is also produced via $^{129}$I, so an elevated I/Xe ratio is implicated. Gilmour & Turner (2007) pointed out that chemical separation can be responsible for large variations of the elemental I/Xe ratio on the local scale. Because of this unresolved radiogenic effect, the case of $^{129}$Xe is not further discussed here.

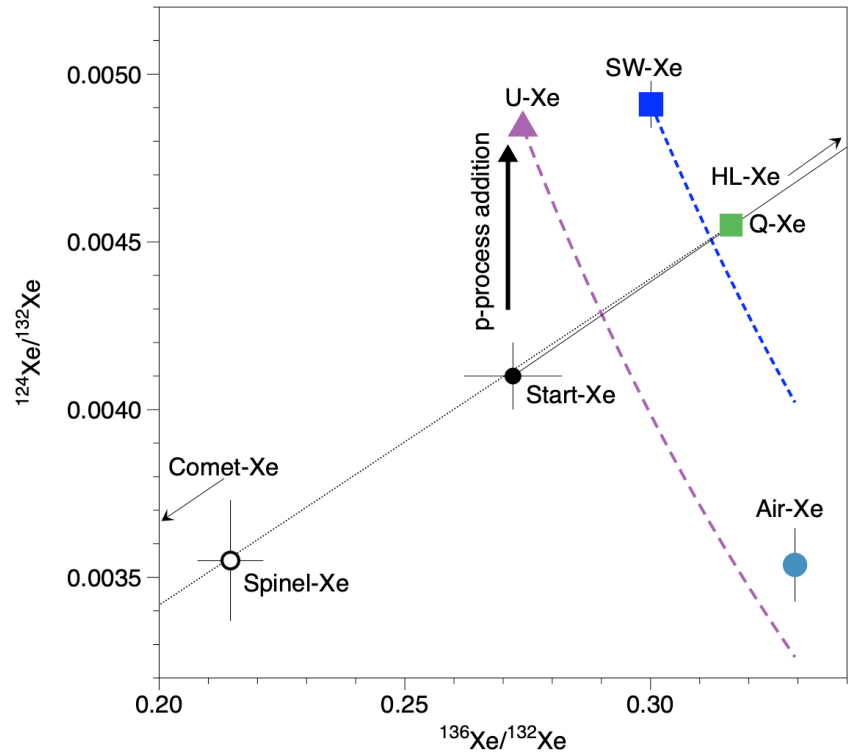

**Figure 5: Three-isotope diagram of Xe components discussed in this study.** The purple and blue dashed-lines represents mass-dependent fractionation of U-Xe and SW-Xe, respectively. Only fractionation of U-Xe leads to Air-Xe (after minor addition of Pu-derived $^{136}$Xe; Pepin 1991). Extrapolation of the nucleosynthetic mix for comets of Marty et al. (2017) and mixing (dotted line) with 78% of Q-Xe leads to a progenitor of Air-Xe depleted in p-process $^{124}$Xe and $^{126}$Xe (only $^{124}$Xe is visible here). Mixing with HL-Xe (black plain line) does not compensate this effect since it would be accompanied by r-process $^{136}$Xe. Errors are at 1σ.

Several processes could be envisaged to explain the depletion in $^{124-128}$Xe of Start-Xe relative to U-Xe and, by extension, to SW-Xe. Corrections of mass-dependent fractionation of Air-Xe and of minor addition of plutonium derived fission Xe led to the definition of U-Xe (Pepin 1991; Fig. 5). While Start-Xe closely resembles U-Xe for $^{129-136}$Xe, the low $^{124-128}$Xe/$^{132}$Xe ratios are incompatible with Start-Xe being a good candidate for starting Earth (or Mars) atmospheric Xe. Mixing HL-Xe with Start-Xe compensates for the depletion in the light isotopes but this component also carries excesses in $^{134}$Xe and $^{136}$Xe (Ott 2014) and would thus lead to a departure from U-Xe for heavy isotopes. Xenon from spallation is an interesting candidate for compensating this depletion since it especially enriched in light Xe isotopes. However, the exposure age or the concentration of target elements would have to be extremely high and the relative proportions of $^{124}$Xe, $^{126}$Xe and $^{128}$Xe for spallation Xe (Wieler 2002) do not match the depletion observed here. The most plausible explanation for compensating the depletion in $^{124-128}$Xe of Comet-Xe (and Start-Xe) relative to SW-Xe is simply the incorporation of pure p-process Xe to the R/S-process mixture considered here. Either the incorporation of p-process Xe from the interstellar medium happened before the incorporation of r-process Xe or material in the outer regions of the solar system do not carry the same mix of presolar sources as what have been preserved in meteorite parent bodies. Incorporation of p-process Xe with $^{124}$Xe/$^{126}$Xe and $^{128}$Xe/$^{126}$Xe ratios of 2.15±0.82 and <12 (Fig. 6), respectively, to Comet-Xe before mixing with Q-Xe would solve the depletion in $^{124-128}$Xe. Interestingly, Ballad et al. (1979) deduced from series of chemical attacks that spinels in the Allende meteorite also carry Xe with an isotopic composition depleted in p- and r-process Xe isotopes (Fig. 3 & 5). The p-process required to bring $^{124-128}$Xe/$^{132}$Xe ratios back to solar values is in the range of the one required in the case of Comet-Xe (Fig. 6). Although results by Ballad et al. (1979) are not direct measurements but a mathematical derivation by comparing Xe isotope ratios before and after a chemical attack by $H_3PO_4$ and $H_2SO_4$, an isotopic pattern similar to Comet-Xe for light isotopes favors the hypothesis that CC-like material originating from the outer solar system carries a nucleosynthetic Xe component similar to what was measured in the coma of 67P/C-G. The p-process isotopic ratios required to reach solar ratios are 2.15 ± 0.82 and a range from 0 to 12 for $^{124}$Xe/$^{126}$Xe and $^{128}$Xe/$^{126}$Xe, respectively. They are compared in Fig. 6 to theoretical ratios derived from six nucleosynthesis models compiled and detailed by Arnould & Goriely (2003) (see refs. therein). Among the six nucleosynthesis models considered, the ones involving the thermonuclear explosion of a CO White Dwarf star with a s-process seed distribution in the white dwarf of one hundred times the solar composition (Goriely et al. 2002) and the one invoking a type 2 supernova of 25 times the solar mass with a p-process layer (PPL) being contaminated by the overlying exploding He envelope do not lead to $^{124}$Xe/$^{126}$Xe and $^{128}$Xe/$^{126}$Xe ratios compatible to those derived in this study. Unfortunately, the absence of measurement for the two rarest Xe isotopes $^{124}$Xe and $^{126}$Xe emitted by 67P/C-G prevent us to further constrain which type of nucleosynthetic event contributed light Xe isotopes to the solar system. In summary, the composition of Comet-Xe derived in this study implies the incorporation of p-process Xe to the original mixture of r and s process Xe to reach solar-like isotopic ratios for $^{124-128}$Xe isotopes. The combination of p-process and r-process Xe for Comet-Xe and Solar-Xe (and by extension chondritic Xe) has to be different in order to preserve the depletion in $^{134}$Xe and $^{136}$Xe, the principal feature of U-Xe, the progenitor of atmospheric Xe. These variations of the contribution of p-process Xe to solar system reservoirs create non mass-dependent variations of Xe isotope ratios and may explain how the study of light Xe isotopes allowed to determine that Xe in the Earth's mantle has a chondritic origin (*e.g.* Péron & Moreira 2018).

Recent studies show that CC-like material from the outer solar system is enriched in neutron-rich (heavy) isotopes of Cr, Ni, Mo etc. These studies point toward an enrichment of r-process nuclides for material in the outer solar system. This seems to conflict with cometary Xe being depleted in r-process isotopes. However, the processes responsible for the preservation of nucleosynthetic anomalies in phases containing noble gases and those carrying more refractory elements are likely different. Cometary xenon likely resides in condensed phases of volatile elements (ices or clathrates) (Altwegg et al. 2015) while the isotopic composition for more refractory elements in meteorites could be controlled by the preservation of different populations of presolar grains (Trinquier et al. 2009). Furthermore, results on Xe for 67P/Churyumov-Gerasimenko only concerns xenon emitted by gas emanations from the comet. The isotopic composition of xenon in the "rocky" part remains unknown and could show a complementary isotopic pattern similar to results obtained by Ballad et al. (1979), for example.

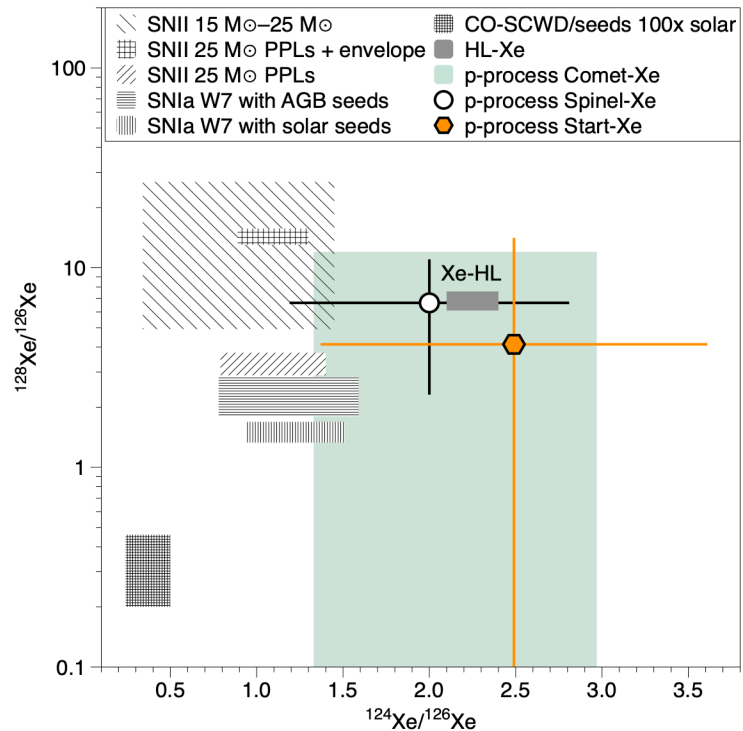

**Figure 6: Three-isotope diagram of Xe for different p-process nucleosyntheses discussed in this study.** The ranges for different models of p-process are from Arnould & Goriely (2003). "PPLs" stands for p-process layers. p-process Comet-Xe (green range) reflects the isotope ratios of pure p-process Xe required to bring $^{124-128}$Xe/$^{132}$Xe ratios back to solar values. Same for p-process

Spinel-Xe (empty circle) and p-process Start-Xe. Error bars and error ranges are at 1σ.

Xenon is not the only element on Earth to show nucleosynthetic departures from the base solar composition or from cosmochemical end-members sampled by chondrites. For example, the silicate Earth shows enrichments in p-process (Andreasen & Sharma 2006) and s-process (Burkhardt et al. 2016) neodymium relative to chondrites. Terrestrial Calcium (Dauphas et al. 2014), Ti (Trinquier et al. 2009), Mo (Budde et al. 2019; Burkhardt et al. 2011) etc. are also showing nucleosynthetic departures relative to chondritic meteorites. Altogether, these results have been used to suggest the persistence of heterogeneities in the solar nebula (Andreasen & Sharma 2006) and the existence of an isotopic reservoir in the inner solar system sampled by the Earth-Moon couple but also by the enstatite chondrites (Dauphas 2017; Dauphas et al. 2014). Recent results obtained on material emitted by 67P/Churyumov-Gerasimenko helped to understand which type of material accreted outside the orbits of the giant planets (Hoppe et al. 2018). However, the nucleosynthetic inventory of the outer solar system remains to be conducted and compared with recent models linking nucleosynthetic anomalies and the formation of the protoplanetary disk (Jacquet et al. 2019). A recent study revealed that a carbon-rich clast found in the CR2 chondrite LaPaz Icefield 02342 is derived from cometary material (Nittler et al. 2019). If more analog samples were found in the future, precise measurements of the concentration and isotopic composition of noble gases in such material with state-of-the-art techniques would complete our understanding of the distribution of volatile elements in the solar system.

## 5. CONCLUSIONS

The precursor of atmospheric Xe presents selective depletions in $^{134}$Xe and $^{136}$Xe (Pepin 1991) suggesting a contribution from a cosmochemical source different from solar gas or meteorites and depleted in r-process Xe isotopes. Measurements of Xe in the coma of comet 67P/C-G identified such a depletion (Marty et al. 2017). Propagation of the data obtained on 67P/C-G toward light Xe isotopes using correlation planes derived in a previous study (Gilmour & Turner, 2007) leads to a starting composition for the Earth's atmosphere different from U-Xe. Incorporation of a pure p-process end-member is required to obtain solar-like isotopic ratios for $^{124-128}$Xe. Unfortunately, this p-process end-member remains elusive and further studies are required to define which type of nucleosynthesis event contributed p-process isotopes to the solar system.

Xenon in the solar system shows large scale variations of r-process and p-process nuclides which implies either that only the inner solar system was contaminated by r-process xenon, maybe by a single event (Bartos & Marka 2019), or a differential survival of nucleosynthetic carriers. A comet sample return mission, even without long-term storage at cryogenic temperatures, or the finding of relatively pure cometary material in meteorite collection would allow precise measurements of the rarest p-only and p/s-process Xe isotopes in cometary ice and grains of comets. This would help to draw the map of nucleosynthetic heterogeneities in the solar system and understand how presolar material was distributed at the time of solar system formation and during its subsequent evolution.

**Acknowledgments**

Members of the CAGE team (Cosmochimie, astrophysique et géophysique expérimentale) at IPGP are thanked for insightful discussions. G.A. acknowledges funding from the Region Ile-de-France through the DIM-ACAV+ program. J.D.G. acknowledges funding from the Science and Technology Facilities Council (STFC) grant number ST/R000751/1. This study contributes to the IdEx Université de Paris ANR-18-IDEX-0001. This is IPGP contribution #4092.

**Table 1: Xenon components used and derived in this study. All errors at 1σ.**

**Table 1**
Xe components used and derived in this study. All errors at 1σ.

| Component | $^{124}Xe/^{132}Xe$ | ± | $^{126}Xe/^{132}Xe$ | ± | $^{128}Xe/^{132}Xe$ | ± | $^{129}Xe/^{132}Xe$ | ± | $^{130}Xe/^{132}Xe$ | ± | $^{131}Xe/^{132}Xe$ | ± | $^{134}Xe/^{132}Xe$ | ± | $^{136}Xe/^{132}Xe$ | ± | Reference |
|---|---|---|---|---|---|---|---|---|---|---|---|---|---|---|---|---|---|
| Air-Xe | 0.00354 | 0.00001 | 0.00330 | 0.00002 | 0.0714 | 0.0001 | 0.983 | 0.001 | 0.1514 | 0.0001 | 0.7890 | 0.0011 | 0.3879 | 0.0006 | 0.3294 | 0.0004 | Basford et al. (1973) |
| SW-Xe | 0.00491 | 0.00007 | 0.00416 | 0.00009 | 0.0842 | 0.0003 | 1.041 | 0.001 | 0.1650 | 0.0004 | 0.8256 | 0.0012 | 0.3691 | 0.0007 | 0.3001 | 0.0006 | Meshik et al. (2014) |
| U-Xe | 0.00484 | 0.00002 | 0.00419 | 0.00002 | 0.0841 | 0.0001 | 1.040 | 0.001 | 0.1654 | 0.0002 | 0.8262 | 0.0013 | 0.3516 | 0.0007 | 0.2740 | 0.0006 | Pepin (2002) |
| 67P/CG-Xe | n.d. | | n.d. | | 0.0760 | 0.0210 | 1.398 | 0.122 | 0.1840 | 0.0320 | 0.8640 | 0.0900 | 0.2380 | 0.0490 | 0.1160 | 0.0400 | Marty et al. (2017) |
| Spinel-Xe | 0.00355 | 0.00018 | 0.00348 | 0.00024 | 0.0797 | 0.0024 | 1.114 | 0.016 | 0.1655 | 0.0028 | 0.8121 | 0.0112 | 0.3116 | 0.0054 | 0.2145 | 0.0067 | Ballad et al. (1979) |
| Comet-Xe | 0.0026 | 0.0004 | 0.0031 | 0.0003 | 0.0853 | 0.0141 | 0.983 | 0.096 | 0.184 | | 0.773 | 0.064 | 0.239 | 0.030 | 0.116 | | this study |
| Start-Xe | 0.0041 | 0.0001 | 0.0038 | 0.0001 | 0.0829 | 0.0031 | 1.029 | 0.022 | 0.167 | 0.001 | 0.809 | 0.014 | 0.347 | 0.010 | 0.272 | 0.010 | this study |